\documentclass{aa}
\usepackage[dvips]{graphicx}
\usepackage{longtable,lscape}

\begin{document}
 
\title{Short period eclipsing binary candidates identified using SuperWASP}
 
\titlerunning{WASP short period eclipsing binaries}
 
\author{A.J.~Norton\inst{1}, S.G.~Payne\inst{1}, T.~Evans\inst{1},
 R.G.~West\inst{2}, P.J.~Wheatley\inst{3}, D.R.~Anderson\inst{4}, 
 S.C.C.~Barros\inst{5}, O.W.~Butters\inst{2},
 A.~Collier~Cameron\inst{6}, D.J.~Christian\inst{5,9}, B.~Enoch\inst{6}, F.~Faedi\inst{5}, 
 C.A.~Haswell\inst{1},  C.~Hellier\inst{4}, S.~Holmes\inst{1}, K.D.~Horne\inst{6}, S.R.~Kane\inst{10},
 T.A.~Lister\inst{7}, 
 P.F.L.~Maxted\inst{4},  N.~Parley\inst{6}, D.~Pollacco\inst{5}, E.K.~Simpson\inst{5}, I.~Skillen\inst{8}, 
 B.~Smalley\inst{4}, J.~Southworth\inst{4}, R.A.~Street\inst{7}} 
 
\authorrunning{Norton et al}
 
\offprints{A.J. Norton, a.j.norton@open.ac.uk}
 
\institute{Department of Physics and Astronomy, The Open University,
           Walton Hall, Milton Keynes MK7 6AA, U.K.
\and
           Department of Physics and Astronomy, University of Leicester,
           Leicester LE1 7RH, U.K.
\and
           Department of Physics, University of Warwick, Coventry CV4 7AL, U.K.
\and
           Astrophysics Group, Keele University, Staffordshire ST5 5BG, U.K.
\and
           Astrophysics Research Centre, Main Physics Building, School of
           Mathematics \& Physics, Queen's University, University Road, 
           Belfast BT7 1NN, U.K.
\and
           School of Physics and Astronomy, University of St. Andrews,
           North Haugh, St. Andrews, Fife KY16 9SS, U.K.
\and 
         Las Cumbres Observatory Global Telescope Network, 6740 Cortona Drive, 
         Suite 102, Goleta, CA 93117, USA
\and
           Isaac Newton Group of Telescopes, Apartado de Correos 321, 
           E-38700 Santa Cruz de la Palma, Tenerife, Spain
\and
            Department of Physics and Astronomy, California State University, 
            Northridge, CA 91330, USA
\and
        NASA Exoplanet Science Institute, Caltech, MS 100-22, 770 South Wilson Avenue, 
        Pasadena, CA  91125, USA}

\date{Accepted ??? 2011;
      Received ??? 2011}
 
\abstract{We present light curves and periods of 53 candidates for short period eclipsing binary stars identified by SuperWASP. These include 48 newly identified objects with periods $<2\times 10^4$~seconds ($\sim 0.23$~d), as well as the shortest period binary known with main sequence components (GSC$2314-0530 =$ 1SWASP J$022050.85+332047.6$) and four other previously known W UMa stars (although the previously reported periods for two of these four are shown to be incorrect). The period distribution of main sequence contact binaries shows a sharp cut-off at a lower limit of around 0.22~d, but until now, very few systems were known close to this limit. These new candidates will therefore be important for understanding the evolution of low mass stars and to allow investigation of the cause of the period cut-off.

}

\maketitle

\section{Introduction}

The primary aim of the SuperWASP photometric survey (Pollacco et al. 2006) 
is to search for transiting extrasolar planets. However, its wide field of view 
(two sets of eight cameras, each of which covers $7.8^{\circ}\times 7.8^{\circ}$) and its 
long base line (operating since 2004) mean that it is also well suited to 
discovering and characterising many types of stellar variability, across (almost) the whole
sky (Norton et al. 2007, Payne et al. 2011). A SuperWASP count rate of 1 count second$^{-1}$ 
is approximately equal to a $V$ magnitude of 15 and this marks the faint limit of the survey. 
At the current time (December 2010), the SuperWASP archive\footnote{http://www.wasp.le.ac.uk/public/} (Butters et al 2010) contains around 275  billion data points on about 30 million unique objects, 
obtained from over 8 million images. The typical observing cadence also means that SuperWASP is 
sensitive to variability periods from around tens of minutes to months.

SuperWASP is thus able to identify many hundreds of thousands of eclipsing binaries,
pulsating stars, and stars displaying rotational modulations of all kinds. The particular
focus of this paper is short period ($P<0.23$~d) eclipsing binary stars. Main sequence stars 
in such close orbits will appear as W Ursa Majoris systems.

Low mass dwarf stars are very common, but the way in which they evolve in close binary systems is poorly 
understood. It has been clear for a number of years that there is a short period cut-off for such 
systems around an orbital period of $\sim 0.22$~d (Rucinski 2007). The reason for this period cut-off was originally thought to be due to stars reaching their fully convective limit (Rucinski 1992), but 
is now believed to be related to magnetic wind-driven angular momentum loss mechanisms, and hence 
linked to the finite age of the binary population (Stepien 2006), although details of the 
process are unclear.

The shortest period binary known with main sequence components is GSC$2314-0530$ which we identified 
using SuperWASP as having a period of 0.1926~d (Norton et al. 2007) and was shown to be coincident 
with a ROSAT X-ray source (1SWASP~J$022050.85+332047.6 = $1RXS~J$022050.7+332049$). This object was
subsequently modelled by Dimitrov \& Kjurkchieva (2010) using multi-colour photometry and 
radial velocity spectroscopy, who showed that it consisted of two stars of mass $0.51~M_\odot$
and $0.26~M_\odot$. Although the two stars are close, the best-fit model has the two stars 
under-filling their Roche lobes, and so the binary is detached. 

The aim of the current work is to use the wide-field capabilities of the SuperWASP survey to 
search for short period W UMa star candidates, in order to increase the sample of such objects substantially and allow detailed investigation of their properties.

\section{Creation of the sample}

A purpose-written period searching code has been run on the entire SuperWASP
archive. As explained in Norton et al. (2007),  the period search comprised 
two techniques: a {\sc clean}ed power spectrum (Lehto 1997) and a folding
technique using $\chi^2$ tests of phase-dispersion minimisation. Each light curve 
with more than 1000 data points was analysed for each object on a per camera basis 
(to minimise the effect of systematic differences between different cameras), and multiple 
significant periods (in some cases) were identified for each. The SuperWASP input catalogue is 
based on the USNO-B1.0 sample of over 1 billion stars, which extends down to magnitude $V \sim 21$.
However, because the SuperWASP pixels are large (around $14^{\prime\prime}$) it is sometimes 
the case that multiple SuperWASP object apertures (each based on the position of a 
single USNO-B1.0 star) sample the variability from a single object. This can result in multiple,
closely spaced SuperWASP objects displaying the same periodic signal. Also, despite using the 
SysRem algorithm (Tamuz et al. 2005) to reduce systematic contaminants in the light curves, 
it is apparent that many spurious periodic signals remain, particularly at harmonics of 1~sidereal day 
(i.e. at periods of 1~sidereal day/$n$ where $n=2,3,4...20,$ etc.). Period ranges within about a minute
either side of these harmonics were therefore flagged as likely spurious in the analysis.

This exercise yielded around 5,600 periodic signals in the period range from about 
125 minutes to 167 minutes (0.087~d to 0.116~d), excluding those period ranges close to 
1/9~d ($\sim 160$~minutes), 1/10~d ($\sim 144$~minutes) and 1/11~d ($\sim 131$~minutes). 
A further 17,300 periodic signals were identified close to 1/9~d; 14,500 signals close to 
1/10~d; and 9,300 signals close to 1/11d. The distribution of the number of objects 
as a function of period within this range is shown in Figure 1. Note that 
a single variable object may give rise to multiple signals if it is observed by multiple 
cameras, or if the variability is sampled by multiple apertures corresponding to closely 
separated stars in the input catalogue. Hence there are fewer unique periodic objects in
this period range than indicated by the numbers quoted above.

Using the period search method described, contact binaries will generally give an identified 
period that corresponds to \emph{half} the orbital period of the system. Hence, the period 
range chosen above covers a range from about 250~minutes (0.175~d) to 333~minutes (0.23~d) 
for eclipsing binary orbital periods. This is therefore the period range of interest, spanning from 
just longer than the suggested period cut-off (0.22~d) to below the period of the shortest known
system (0.19~d). 

\section{Results}

Each of the 5,600 light curves with periods outside the three `contaminated' period regions were 
examined by eye to pick out candidate eclipsing binaries. Most of these light curves displayed 
broadly sinusoidal modulation so are likely to represent some form of rotational variability.
Many others showed asymmetric light curves (with narrow maxima and broad minima) characteristic 
of pulsating stars. These objects will be reported elsewhere. 40 unique objects, reported here, 
displayed light curves that are characteristic of W UMa stars, namely broad maxima 
with narrow minima (see Figure 2, Figure 3 and Table 1). 

The large number of apparently periodic light curves close to periods of 1/9~d, 1/10~d and 1/11~d
were not all eye-balled, as the vast majority of these are expected to be spurious signals due to 
systematic noise. However, the period distribution in Figure 1 shows that $\sim 1\%$ of these are 
likely to be genuine, so the $\sim 1\%$ of them showing the strongest signals \emph{were}
examined individually. The strongest signals were chosen because the signal we are looking for is 
generally of higher amplitude than our typical systematic noise signal. This allowed a further 13 candidate
eclipsing systems to be identified in the set of objects with periods close to 1/9~d 
(see Figure 2 and Table 1). This exercise also revealed the known sdB+dM eclipsing binary, NY~Vir (see 
Figure 4), with a period of 145.463~minutes, within the set of objects with periods close to 1/10~d. No 
candidate W UMa stars were found in this latter set or in the set of objects close to periods of 1/11~d, 
but the clear detection of NY~Vir shows that such binaries would have been found, had 
they existed, despite the systematic noise close to these periods. Figure 5 shows the orbital period 
distribution of the 53 candidate eclipsing binaries found and listed in Table 1.

Further evidence for the fact that these candidates are all W UMa stars is provided by their $V-K$ 
colours which are each significantly redder than the majority of the rest of the objects detected in 
this period range, as shown in Figure 6. In fact, their $V-K$ colours of $\sim 1.5 - 3.5$ are typical of stars of spectral type K. 

The 53 candidate eclipsing binaries include the known W UMa stars: V1067 Her, V1104 Her, ROTSE1 J170240.11+151122.7 and ASAS J111932--3950.8, as well as the shortest period object previously known, GSC$2314-0530$, described earlier (Dimitrov \& Kjurkchieva 2010). The SuperWASP light curve of this latter object is shown separately in Figure 3, folded at its confirmed period of 0.1926~d. Whilst we agree with the periods previously reported for ASAS J111932--3950.8 (0.2294~d, Parihar et al 2009) and V1104 Her (0.2279~d, Akerlof et al 2000), we find better periods for V1067 Her and ROTSE1 J170240.11+151122.7. The ROTSE survey had previously suggested periods of 0.2581~d and 1.210~d respectively for these two objects (Akerlof et al 2000), but we show them to have periods of 0.2285~d and 0.2312~d respectively.

\section{Discussion}

\subsection{Previous short-period samples}

For a number of years the shortest period main sequence eclipsing binary known was the faint (V = 18.3) object OGLE BW03 V038, with an orbital period of 0.1984~d (Maceroni \& Rucinski 1997; Maceroni \& Montalban 2004). This was modelled as consisting of two M3 dwarfs which are almost, but not quite in contact, with masses of $0.44~M_\odot$ and $0.41~M_\odot$. This object is not present in the SuperWASP database owing to its faintness.

In presenting the short-period end of the contact binary period distribution based on the All-Sky
Automated Survey (ASAS), Rucinski (2007) found only three systems in the period range 0.200~d $< P <$
0.225~d (and none at shorter periods). The only one of these they list as a confirmed W UMa system is ASAS J083128+1953.1 with a period of 0.2178~d. Rucinski \& Pribulla (2008) subsequently presented further photometric and spectroscopic observations of this system and argued the case for this system as the shortest period field contact binary. This object is indeed present in the SuperWASP database (as 1SWASP J083127.87+195303.5), with the same period, but the light curve is poorly sampled (fewer than 1000 data points) and jumps between two brightness levels, presumably as a result of another star sometimes falling within the photometric aperture. It nonetheless displays a broadly sinusoidal modulation when folded at this period (see Figure 7a). Since the modulation profile did not match that of a classic contact binary, it was \emph{not} selected as a candidate eclipsing binary by the current exercise.

More recently, Pribulla, Vanko \& Hambalek (2009) looked at two of the candidate short period systems 
listed by Rucinski (2007). They found that one of these stars (ASAS J113031--0101.9) in fact has a period
longer than that originally suggested (0.2710~d instead of 0.2131~d). The other
candidate (ASAS J071829--0336.7) was confirmed as a contact binary with a period of 0.2113~d. As with the 
object discussed by Rucinksi \& Pribulla, this one too is present in the SuperWASP database (as 1SWASP J071828.67--033639.5), but the light curve is very poorly sampled (fewer than 100 data points). 
Nonetheless, when the SuperWASP data are folded at the ASAS period, a profile characteristic of a W UMa star is revealed (see Figure 7b). However, given the few data points, this object was not included in the sample that were examined here.

A survey of a 0.25 square degree region of the Galactic plane using the ESO-2.2m telescope by Miller et al (2010) yielded more than half a million light curves down to $R \sim 24.5$. Amongst this set they found 1318 variable stars, 533 of which were W UMa stars. This included seven candidate contact binaries with periods less than 0.23~d, three of which have periods at or below the period cut-off: V-737 (0.2109~d), V-301 (0.2143~d) and V-1085 (0.2199~d). Each of these stars is relatively faint ($R$ magnitudes in the range $\sim 19 - 22$) and so they are not amenable to detailed follow-up. None of these are present in the SuperWASP database, owing to their location and faintness. 

As noted in the introduction, all of the stars discussed above have recently been superseded as the shortest period eclipsing binary with main sequence components by GSC 2314--0530 (= 1SWASP J022050.85+332047.6) with a period of 0.1926~d (Norton et al 2007; Dimitrov \& Kjurkchieva 2010), as shown in Figure 3 of the present paper.

As a result of our work, however, Figure 5 and Table 1 show that we can now add a further 22 candidates to the set of eclipsing binaries with the shortest periods (0.200~d -- 0.225~d). Four of these are brighter than magnitude 13, so might have been accessible to ASAS. In addition we find a further 30 candidates in the period range at which the sharp cut-off occurs (0.225~d -- 0.230~d), although four of this latter set were already known, two of them had the wrong period previously recorded. Rucinski (2007) commented at the time of the ASAS work that the statistics of the sample at such short periods were very limited, despite the fact that ASAS covered around 3/4 of the sky and extended down to magnitude $\sim 13$. SuperWASP covers a similar fraction of the sky (avoiding the Galactic plane) but is sensitive down to magnitude $\sim 15$, i.e. over six times fainter, and as a result increases the sample by a factor of more than six.

\subsection{Parameter relationships}

Deb \& Singh (2010) have recently presented an analysis of the light curves of 62 binary stars (mostly contact binaries) from the ASAS-3 survey. In particular they show that there is a rather tight correlation between the period and ($J-K$) colour of the contact binaries in their sample (see their Figure 11). The relationship is parameterized by
\begin{equation}
(J-K) = (0.11 \pm 0.01) P^{-1.19 \pm 0.08}
\end{equation}
where $P$ is in days. Although their relationship is well constrained, the majority of their sample have periods around 0.4~d and they include relatively few objects close to the period cut-off. In Figure~8 we show the 2MASS ($J-K$) colours of our sample, against period, with Equation 1 over-plotted. Although the fit is reasonably good, there is significant scatter in the colours of these short period systems.

Gazeas \& Stepien (2008) and Gazeas \& Niarchos (2006) demonstrated that there are clear correlations between the masses of the components of contact binaries and their orbital periods, and also between the radii of the components and their orbital periods. In particular, the period cut-off of around 0.22~d corresponds to primary and secondary masses of around $0.85~M_\odot$ and $0.3~M_\odot$ and radii of around $0.7~M_\odot$ and $0.5~M_\odot$ respectively. These masses and radii correspond to stars of spectral type K and thus match the observed $V-K$ colours we see in our sample (Figure 6). The mass and radius estimates of Gazeas \& Stepien (2008) are good to around 15\% accuracy. However, the short period end of these correlations is defined  by only three stars: CC~Com (0.2211~d), V523~Cas (0.2337~d) and RW~Com (0.2373~d), and there is considerable scatter between their parameters. Increasing the number of objects in this period range which can be modelled will therefore allow these correlations to be tested further and improved upon.

Another interesting result is that of Qian (2001) who demonstrates a possible relation between the period change and the mass ratio of contact binaries. The systems he studies are all at longer periods than those discussed here, so it would be interesting to look for possible period changes in these more compact systems to see if the effect is present here also. In fact, a few of the objects in our samples may already display evidence for period changes. The folded light curves of 1SWASP~J150822.80--054236.9 and 1SWASP~J183738.17+402427.2 for instance, show a broadening of their profiles which cannot be reduced by refining the period further. This is likely to indicate that the period is changing slightly over the several year baseline of the SuperWASP observations.

\subsection{The evolution of low mass binaries}

Low mass binaries are born with a certain minimum period on the main sequence but angular momentum losses, driven by stellar winds, cause such binaries to evolve to even shorter periods as they age. Stars in such systems will expand as their core hydrogen is depleted and reach contact with their Roche lobes within a few billion years, when the orbital period has reduced to around 0.4~d (Stepien, 2006). Detailed calculations by Stepien (2006) show that the time needed to reach a stage of Roche lobe overflow (RLOF), and hence appear as a contact binary, is around 7.5~Gyr for a primary star of $1~M_\odot$. However, because the angular momentum loss timescale increases substantially with decreasing stellar mass, the time to reach RLOF for a system with a primary star of mass $0.7~M_\odot$ increases to greater than the age of the Universe. As a result, Stepien (2006) showed that the short period limit of $\sim 0.22$~d for contact binaries in the Galactic disc corresponds to a lower limit of around $1.1-1.2~M_\odot$ for the total mass of the system (slightly less massive for systems in globular clusters). The short period cut-off for contact binaries is therefore suggested to be due to the finite age of the Galaxy -- lower mass stars which could evolve to shorter periods have not yet had enough time to do so.

Stepien (2006) also notes some limitations of his analysis. In particular, he only performs calculations for initial mass ratios of 0.5 and 1.0 and assumes that the total angular momentum of the binary is approximated by the orbital angular momentum only, which rules out systems with extreme mass ratios. Stepien also notes that if a binary happens to lose a significant fraction of its angular momentum by other mechanisms (such as interactions with a third body, collisions within a dense stellar environment, or enhanced angular momentum loss during the pre-main sequence phase), then the orbital period during the RLOF phase may be as small as 0.15--0.20~d. This may explain the existence of a system such as GSC 2314--0530, with a period of 0.1926~d, which Dimitrov \& Kjurkchieva (2010) modelled as comprising stars of mass $0.51~M_\odot$ and $0.26~M_\odot$. 
\section{Conclusions}

Prior to our work only a handful of W UMa stars at or below the short period cut-off were known. As a result of the candidates presented here, we now have 22 candidates in the period range considered to be below the cut-off (i.e. 0.200~d -- 0.225~d) and a further 30 candidates in the period range of the cut-off itself (0.225~d -- 0.230~d). Along with the intriguing very short period system GSC 2314--0530, there are now well over 50 systems which can be investigated to test the ideas put forward by Stepien (2006) concerning the evolution of low mass binary stars and the mass-radius-period relationships outlined by Gazeas \& Stepien (2008). We urge others to carry out multi-colour photometry and radial velocity spectroscopic follow-up of these systems to confirm their nature.

\begin{acknowledgements}

The WASP project is funded and operated by Queen's University Belfast, the 
Universities of Keele, St. Andrews and Leicester, the Open University, the
Isaac Newton Group, the Instituto de Astrofisica de Canarias, the South
African Astronomical Observatory and by STFC. 

This research has made extensive use of the SIMBAD database, operated at CDS, 
Strasbourg, France. We thank Harry Lehto for use of his implementation of 
the 1D {\sc clean} algorithm.

\end{acknowledgements}

\clearpage

 \begin{figure*}[t]
\begin{center}
\includegraphics[scale=0.5,angle=0]{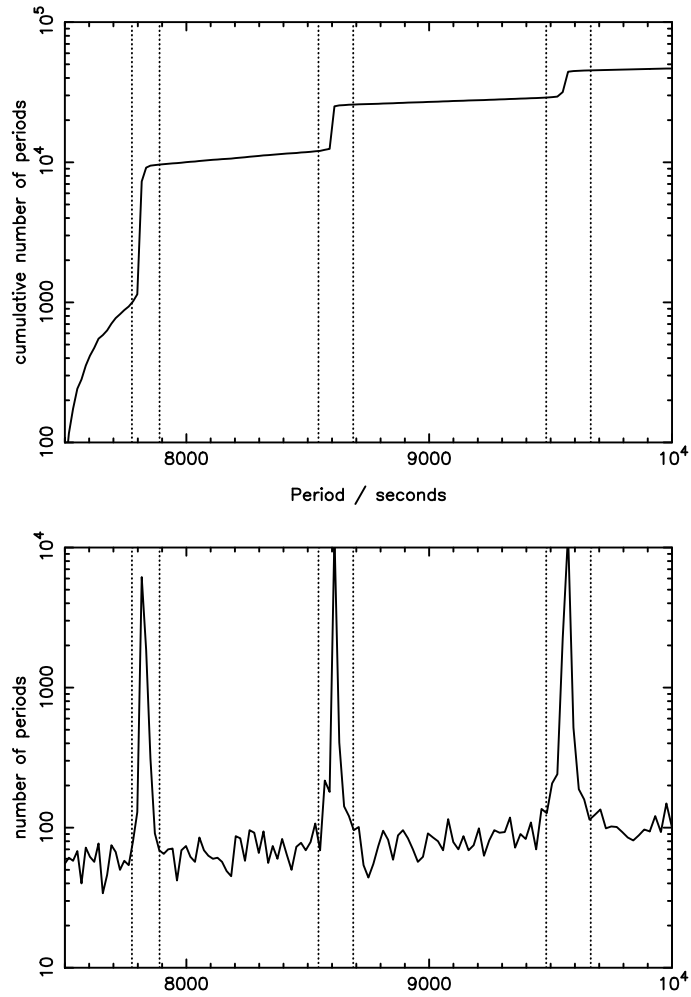}
\caption{The distribution of the periodically variable objects found in the period range from
125 minutes (7,500~seconds) to 167 minutes (10,000~seconds). Note the peaks at periods close to 1/11~d, 1/10~d and 1/9~d which are due to spurious signals caused by systematic noise. Periods within the ranges
indicated by the dotted lines were flagged as likely spurious. Note also that contact binaries
will typically give signals at {\em half} their true orbital period, so the period range here corresponds
to contact binary orbital periods from 250 minutes to 333 minutes.}
\end{center}
 \end{figure*}

\onecolumn

\clearpage

\begin{figure*}[th]
\begin{center}
\includegraphics[scale=0.75,angle=0]{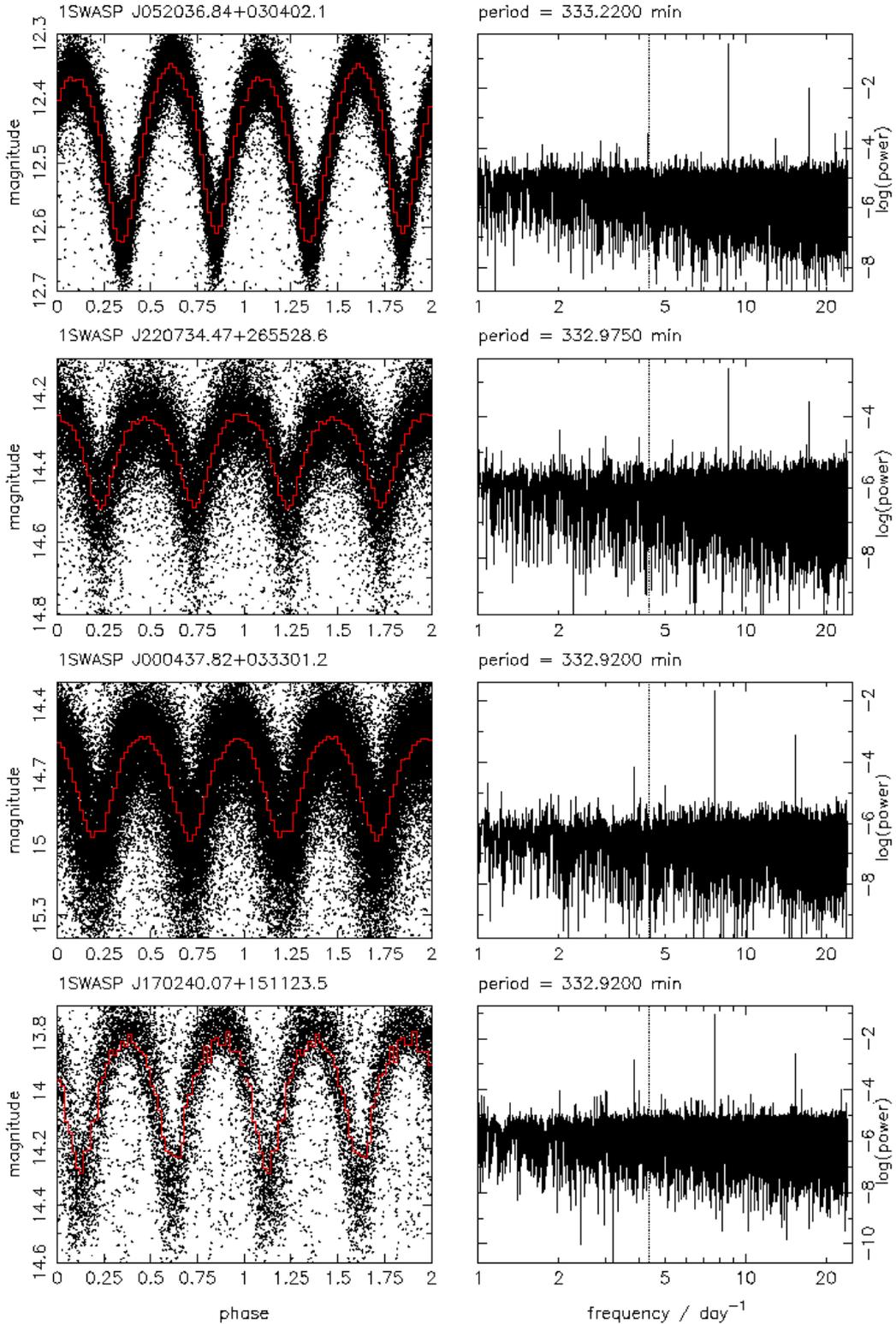}
\caption{(left) SuperWASP light curves folded at the 
orbital period, in order of decreasing period. 
Phase zero corresponds to 2004 January 1st 00:00UT in 
each case. The mean folded light curve (in 50 bins) is shown by an over-plotted 
line. (right) The associated power spectra,
with the frequency corresponding to the presumed orbital period indicated by
a dotted line. The SuperWASP identifier and the presumed orbital period are written
above each pair of panels.}
\end{center}
\end{figure*}

\clearpage
  
\begin{center}
\includegraphics[scale=0.75,angle=0]{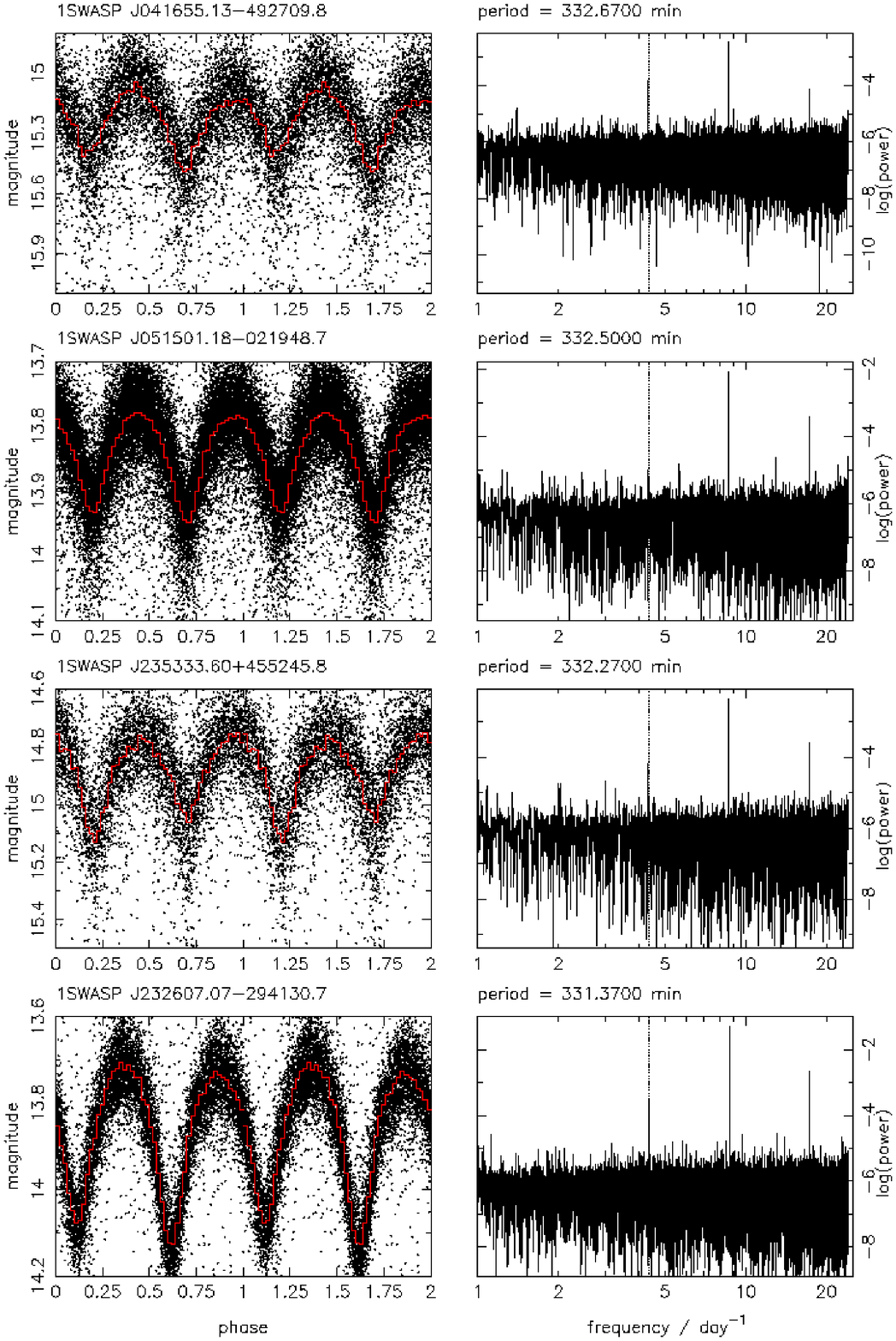}
\end{center}

\begin{center}
\includegraphics[scale=0.75,angle=0]{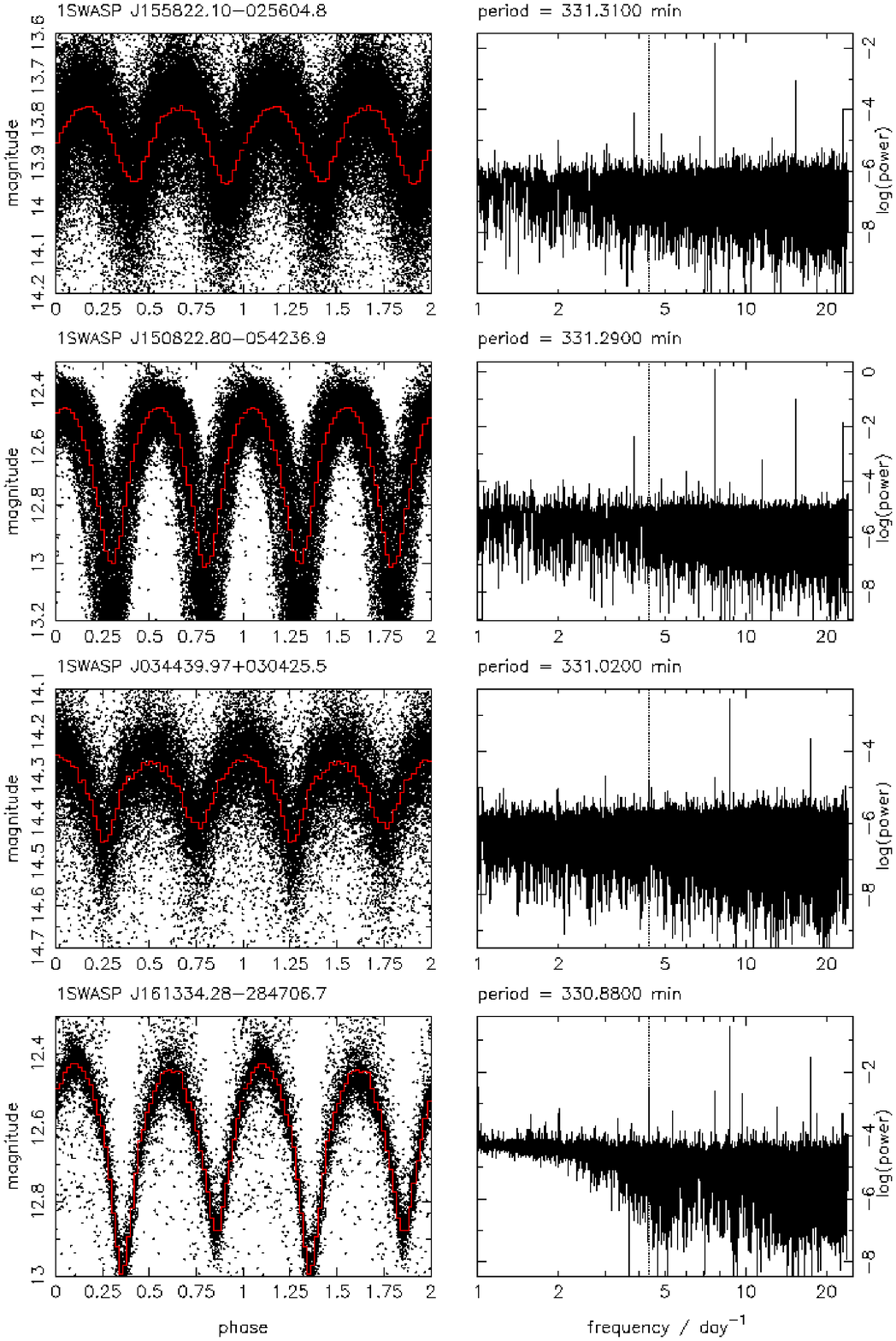}
\end{center}

\begin{center}
\includegraphics[scale=0.75,angle=0]{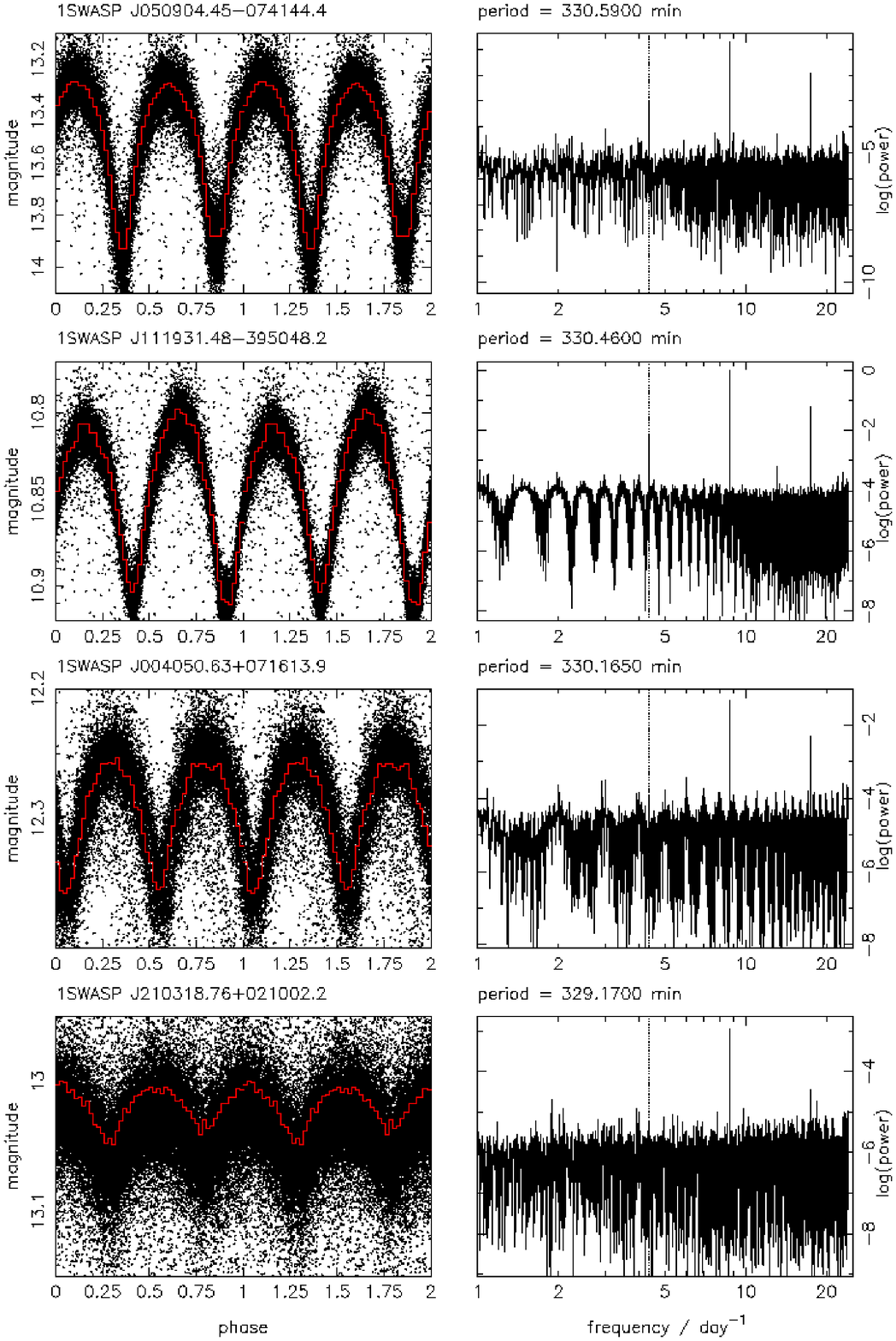}
\end{center}

\begin{center}
\includegraphics[scale=0.75,angle=0]{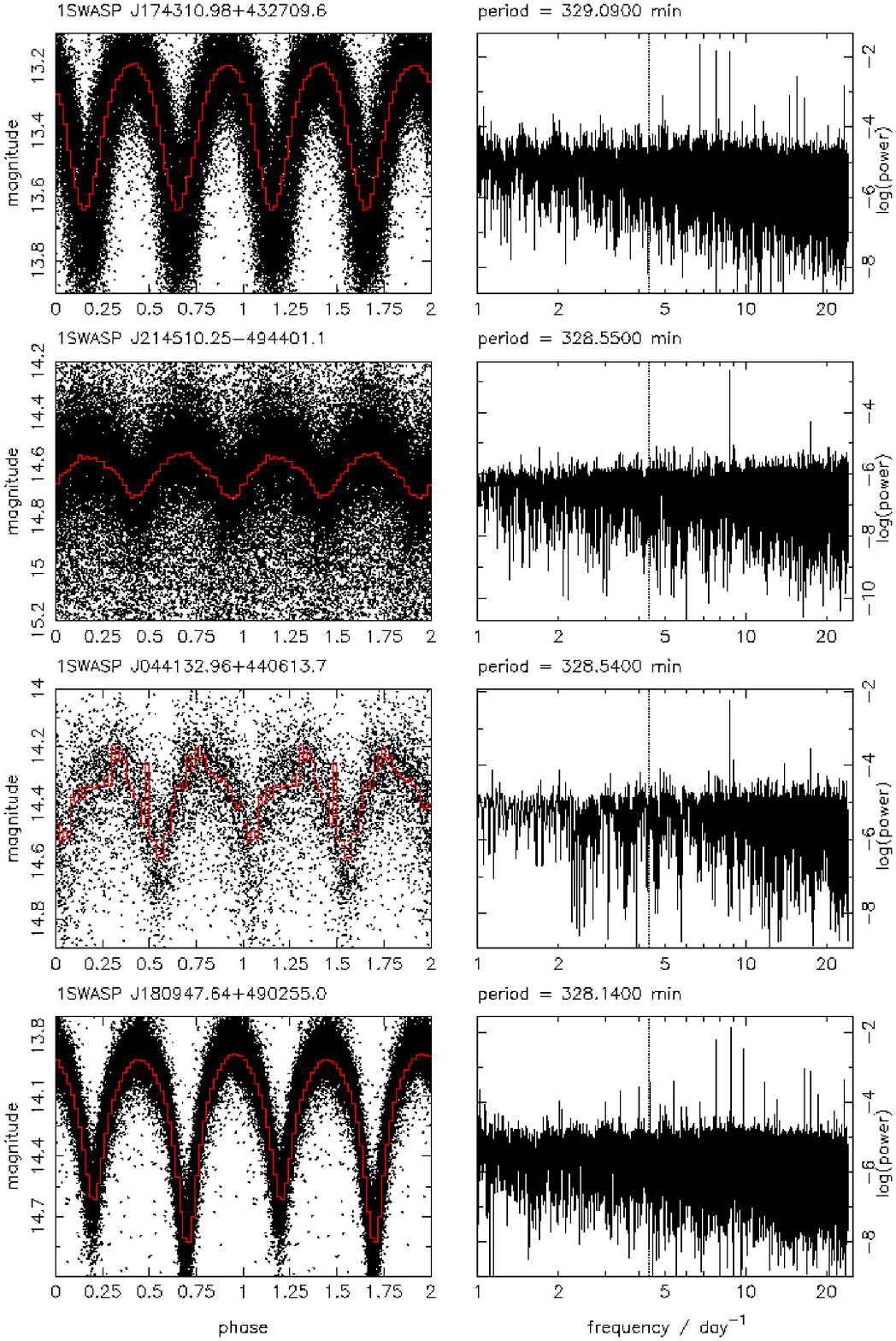}
\end{center}

\begin{center}
\includegraphics[scale=0.75,angle=0]{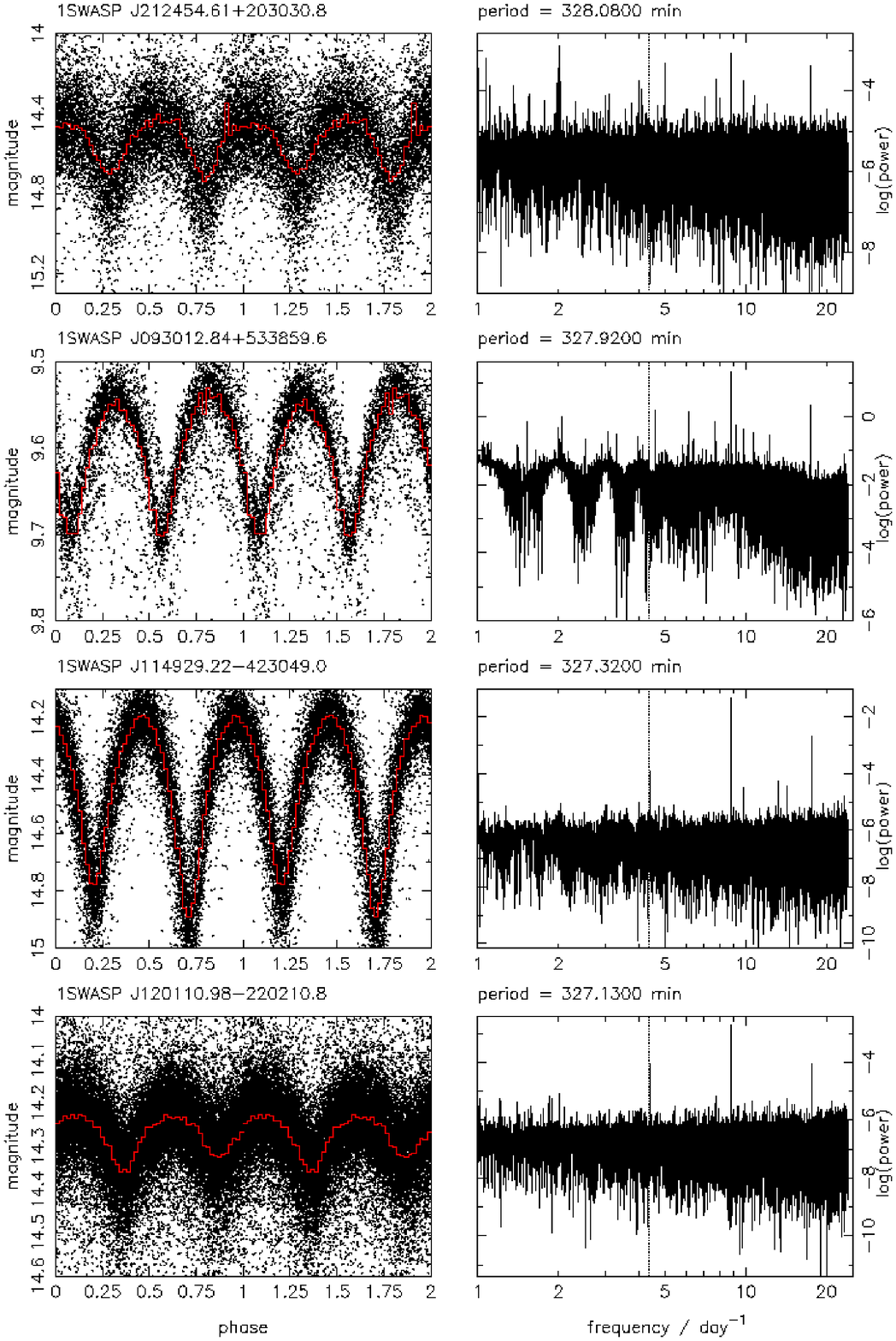}
\end{center}

\begin{center}
\includegraphics[scale=0.75,angle=0]{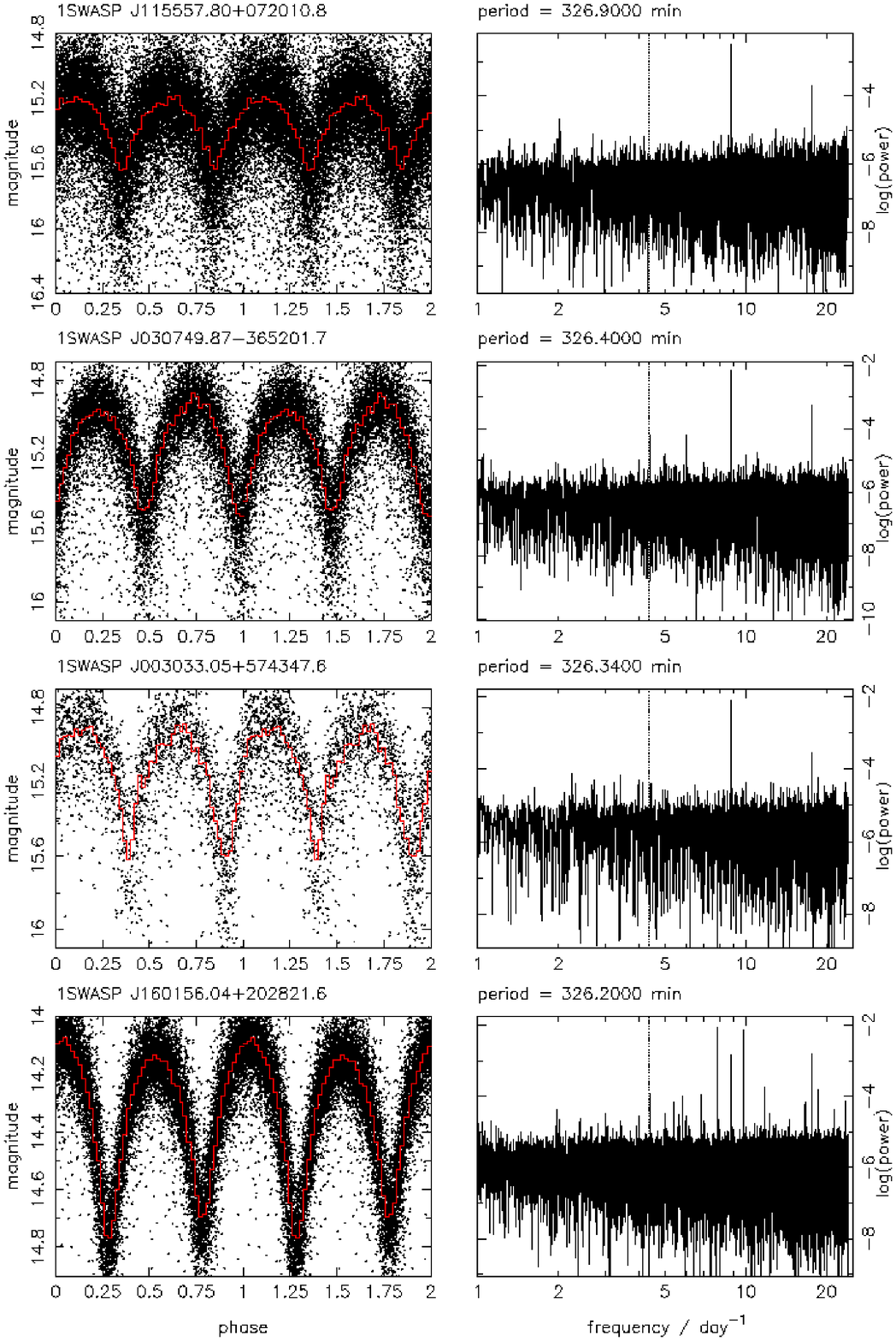}
\end{center}

\begin{center}
\includegraphics[scale=0.75,angle=0]{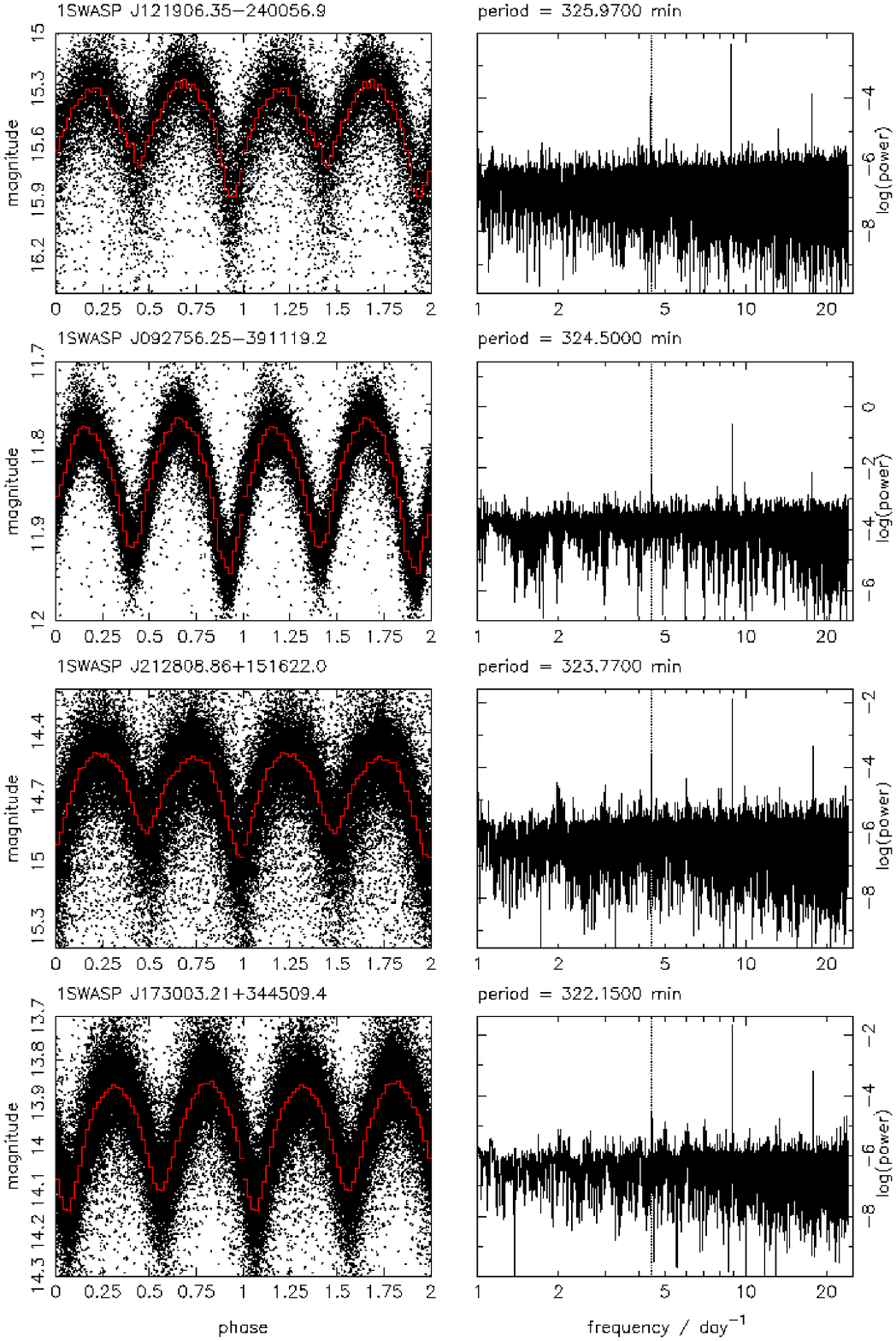}
\end{center}

\begin{center}
\includegraphics[scale=0.75,angle=0]{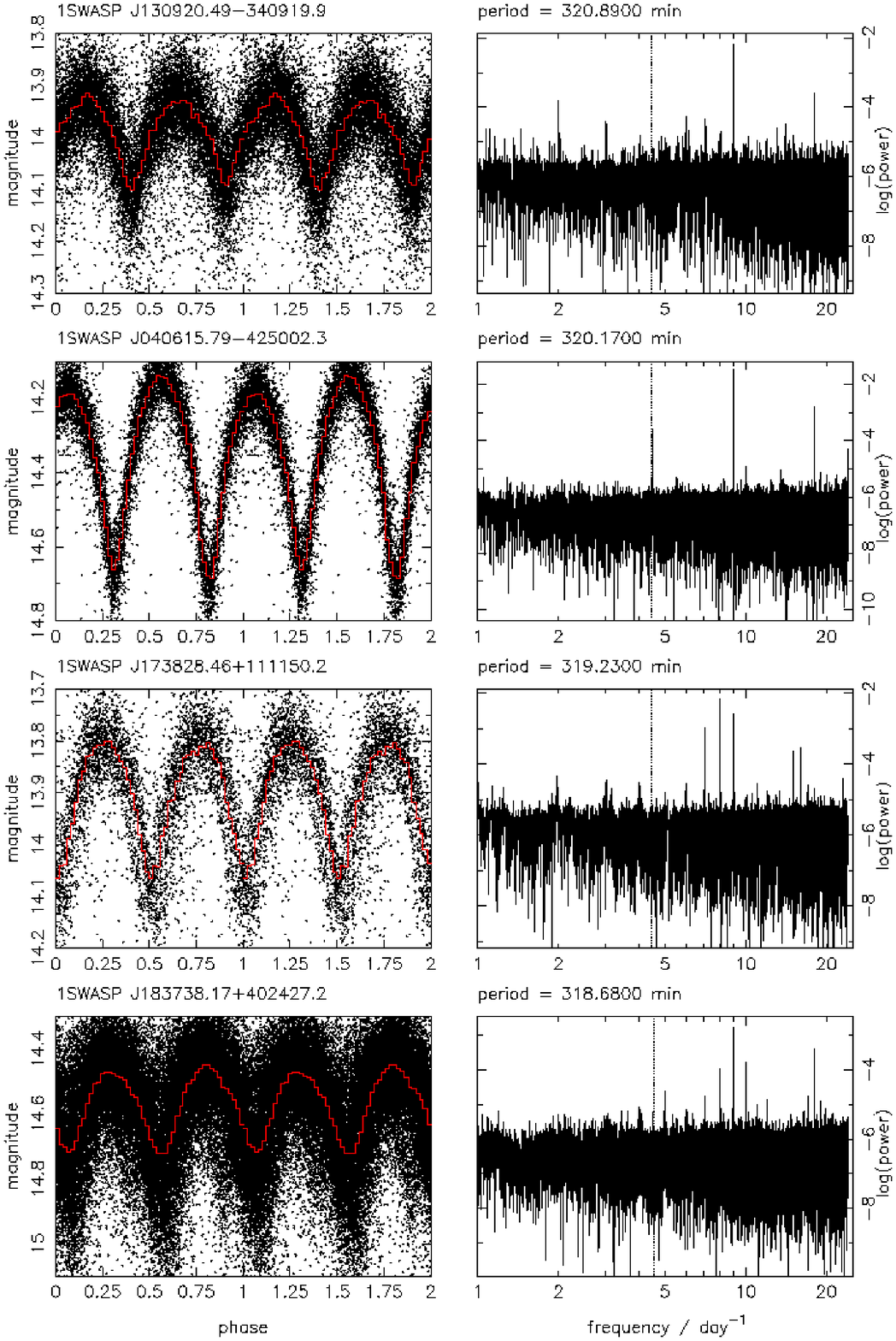}
\end{center}

\begin{center}
\includegraphics[scale=0.75,angle=0]{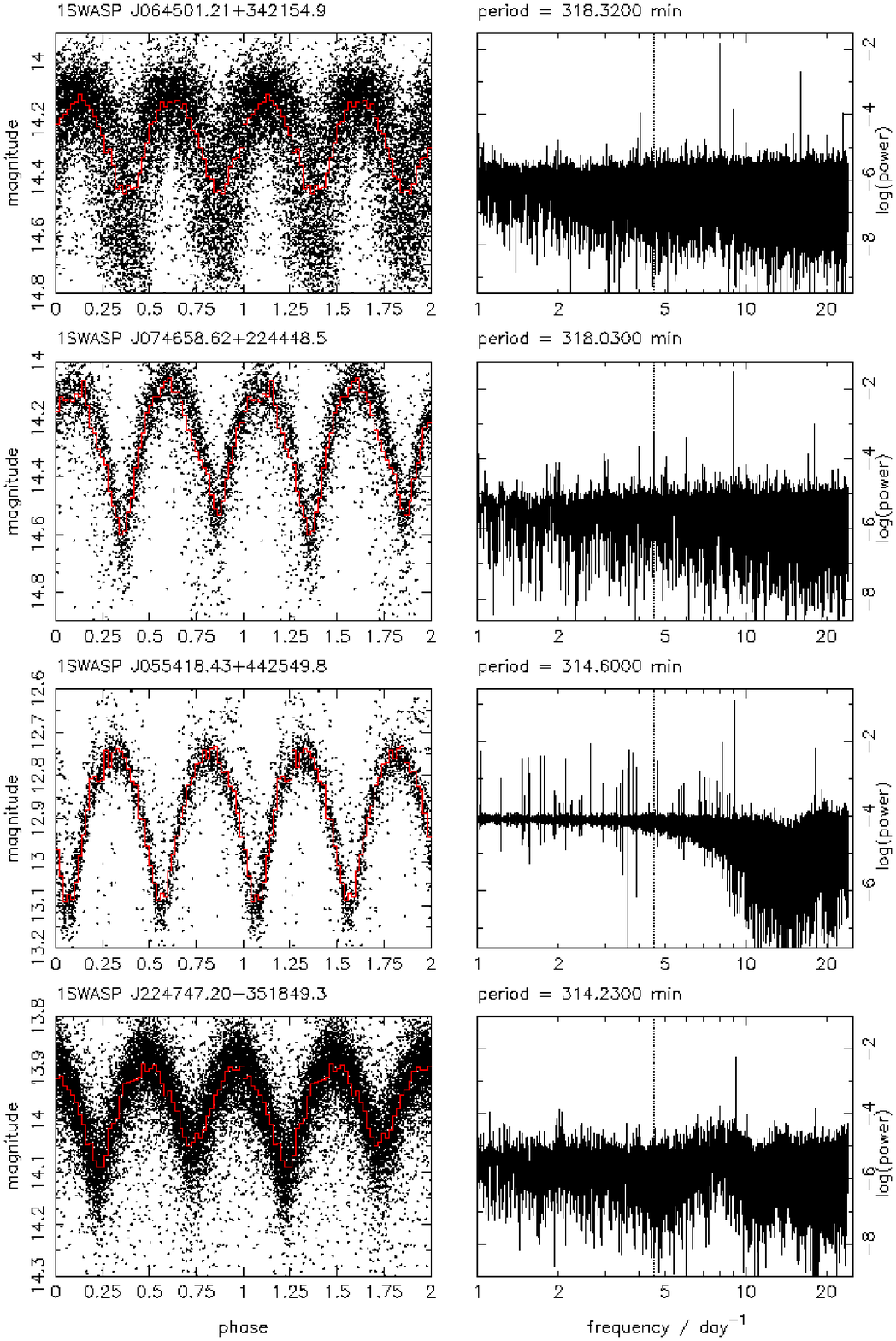}
\end{center}

\begin{center}
\includegraphics[scale=0.75,angle=0]{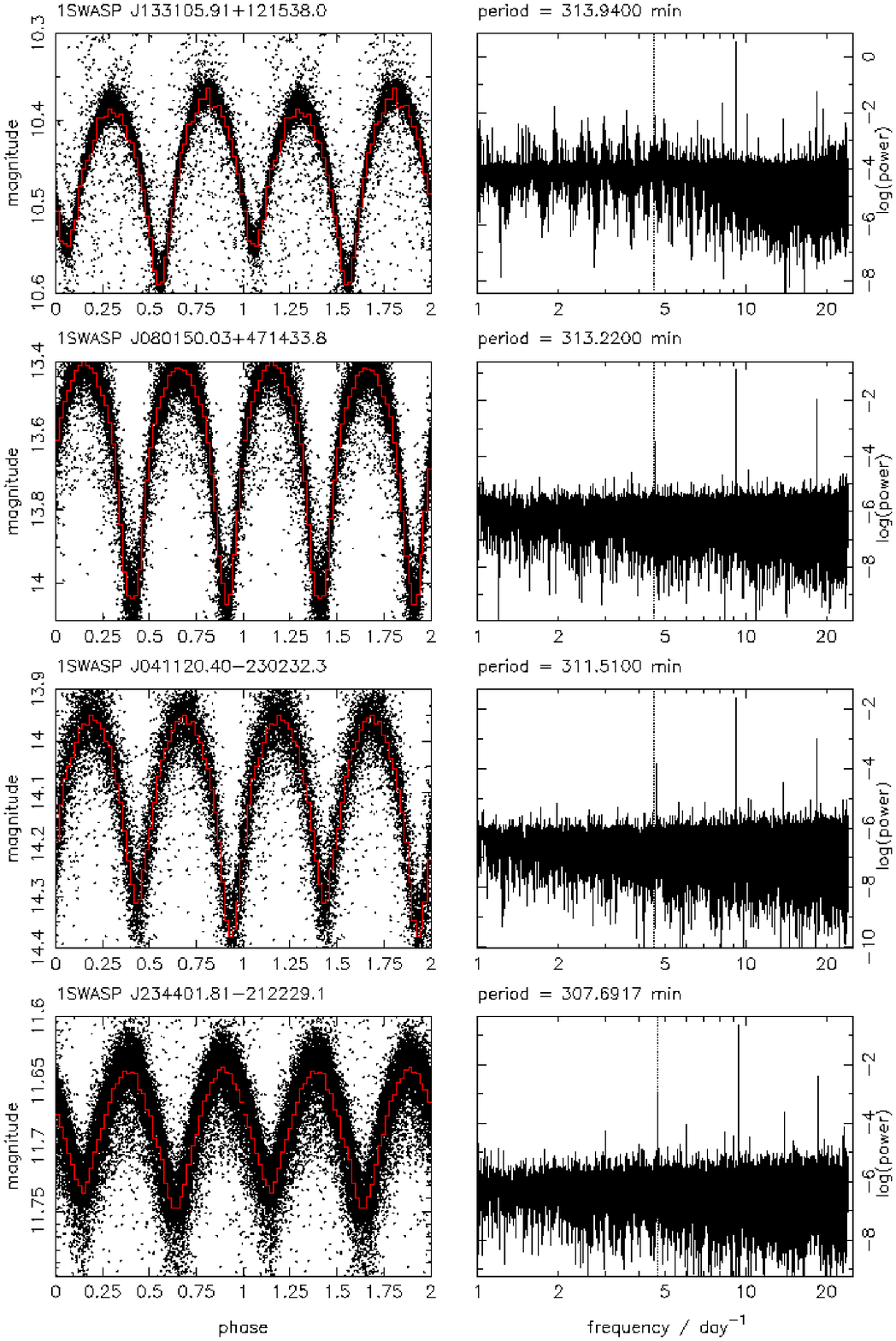}
\end{center}

\begin{center}
\includegraphics[scale=0.75,angle=0]{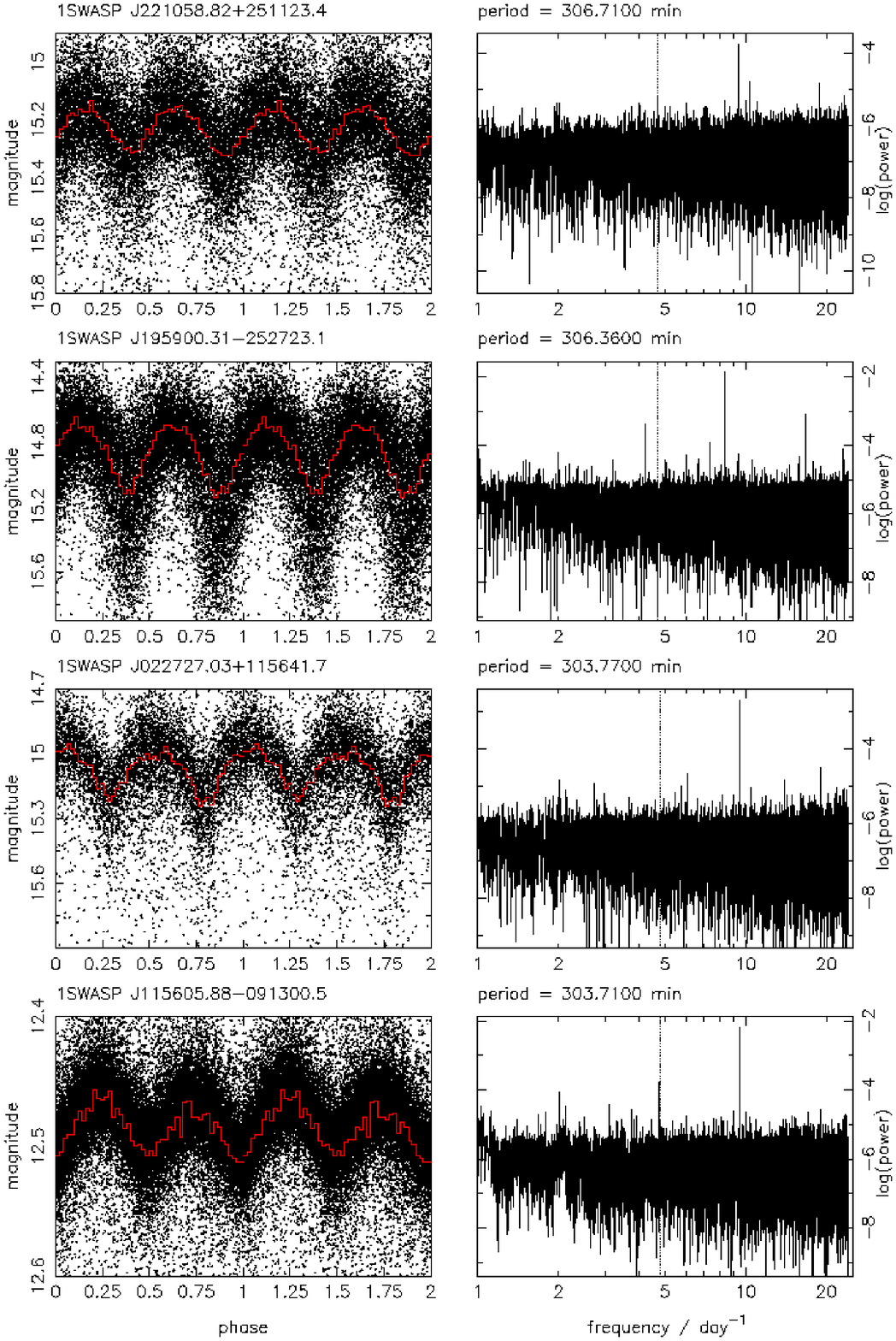}
\end{center}

\begin{center}
\includegraphics[scale=0.75,angle=0]{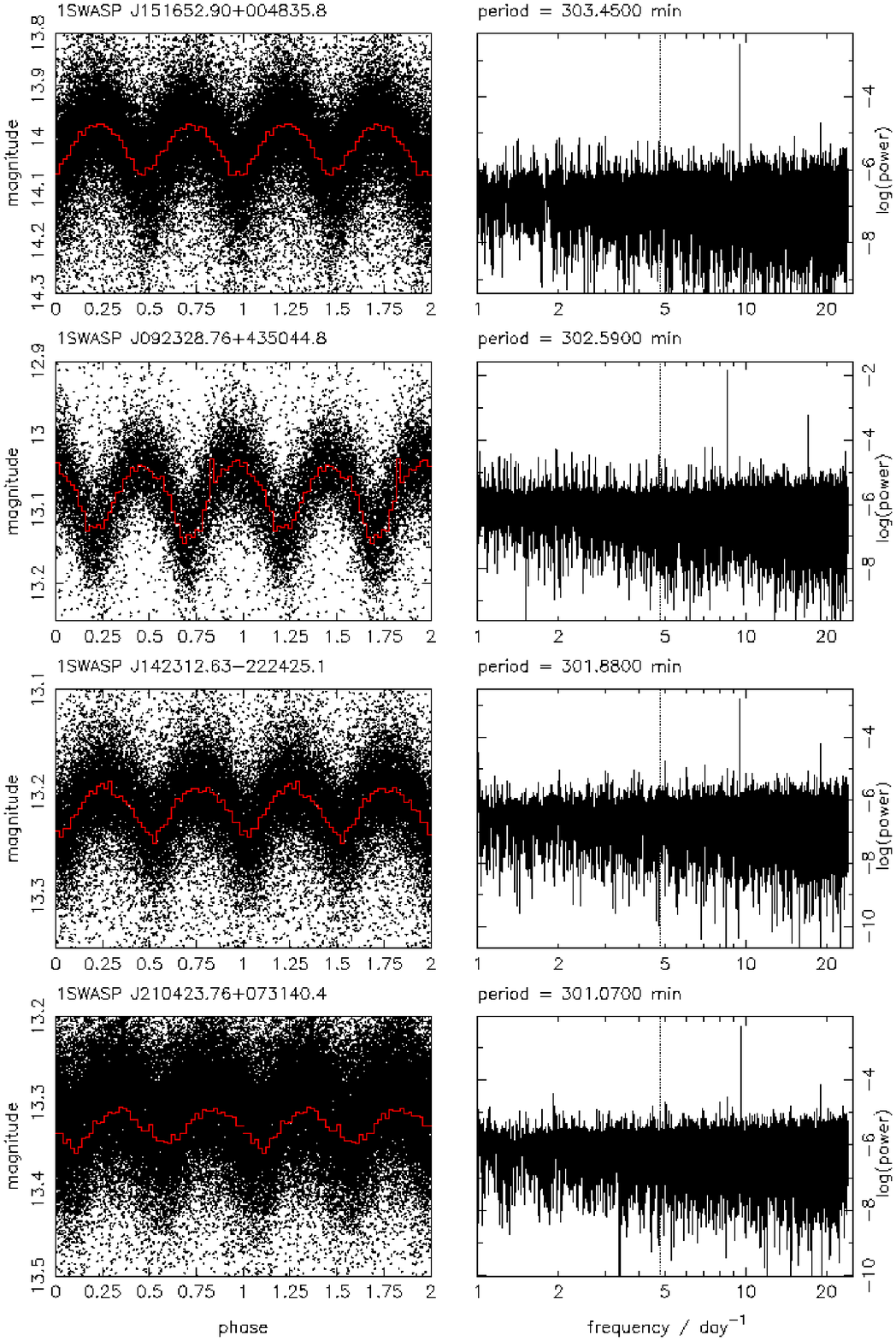}
\end{center}

\pagebreak

\twocolumn

 \begin{figure*}[t]
\begin{center}
\includegraphics[scale=0.75,angle=0]{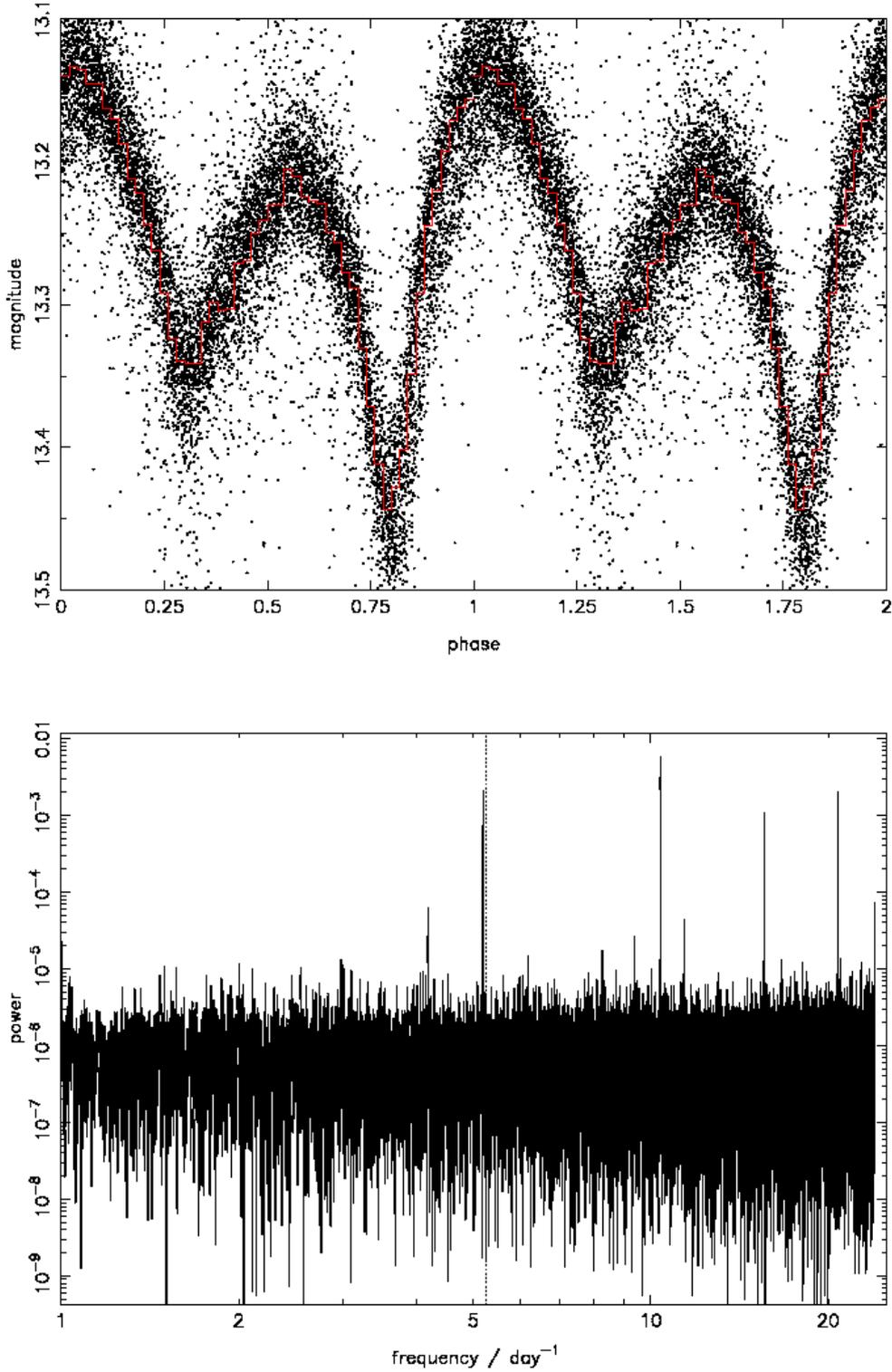}
\caption{(top) The SuperWASP light curve of the shortest period binary known with dM components, 
(GSC$2314-0530 =$ 1SWASP J$022050.85+332047.6$), folded at a period of 277.395 minutes. Phase zero corresponds to 2004 January 1st 00:00UT. The mean folded light curve (in 50 bins) is shown by an over-plotted line. (bottom) The associated power spectrum, with the frequency corresponding to the orbital period indicated by a dotted line. }
\end{center}
 \end{figure*}

 \begin{figure*}[t]
\begin{center}
\includegraphics[scale=0.75,angle=0]{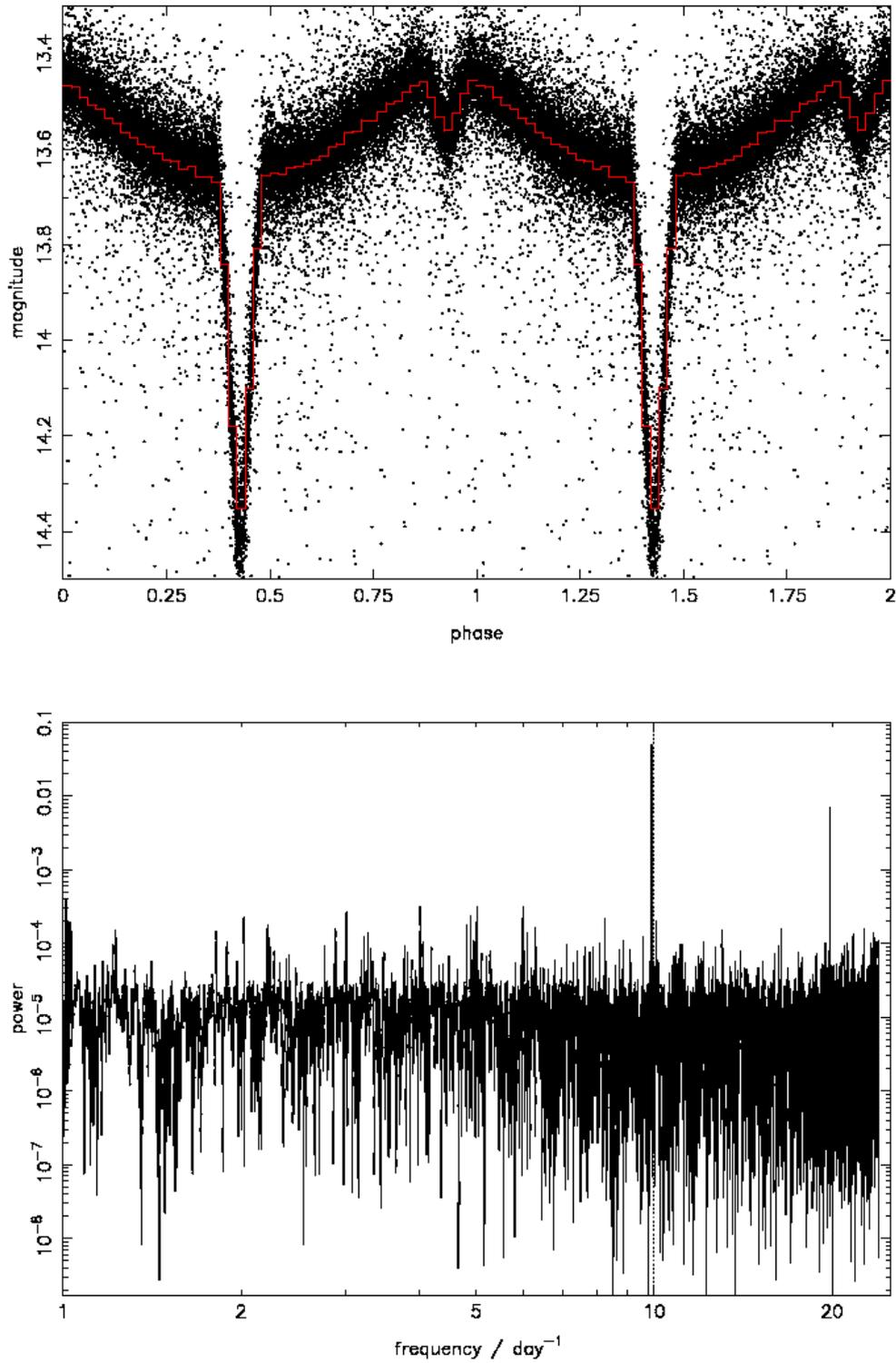}
\caption{(top) The SuperWASP light curve of the sdB+dM eclipsing binary NY~Vir (1SWASP~J$133848.16-020149.3$), folded at a period of 145.463 minutes. Phase zero corresponds to 2004 January 1st 00:00UT. The mean folded light curve (in 50 bins) is shown by an over-plotted line. (bottom) The associated power spectrum, with the frequency corresponding to the orbital period indicated by a dotted line. }
\end{center}
 \end{figure*}

 \begin{figure*}[t]
\begin{center}
\includegraphics[scale=0.5,angle=-90]{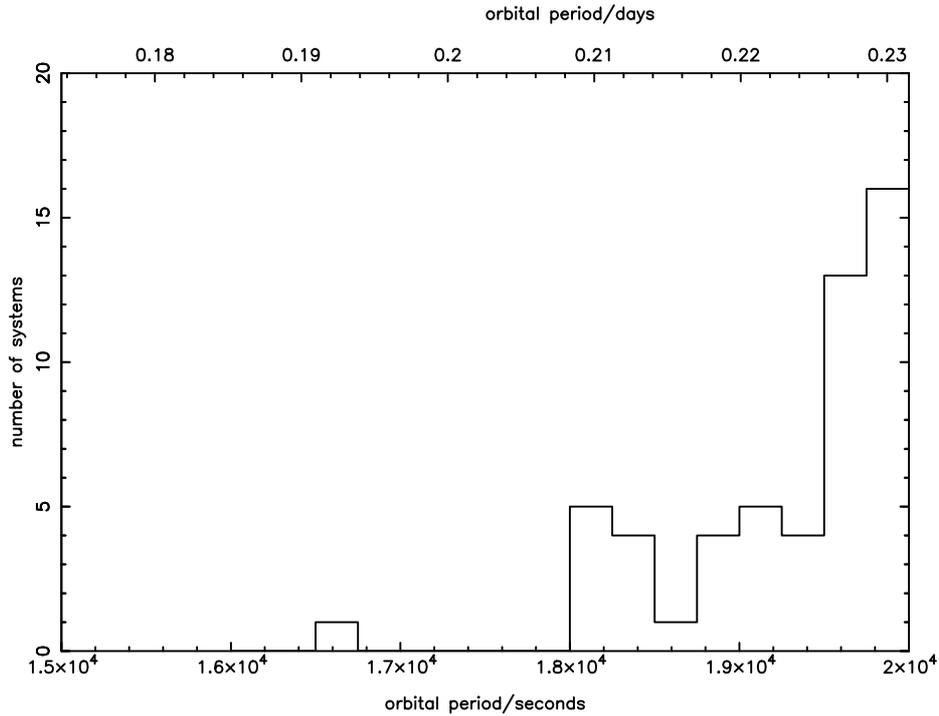}
\caption{The orbital period distribution of the candidate eclipsing binaries found in the range 15,000~seconds ($\sim 0.175$~d) to 20,000~seconds ($\sim 0.23$~d). The bin size is 250 seconds.}
\end{center}
 \end{figure*}
 
 \begin{figure*}[t]
\begin{center}
\includegraphics[scale=0.5,angle=-90]{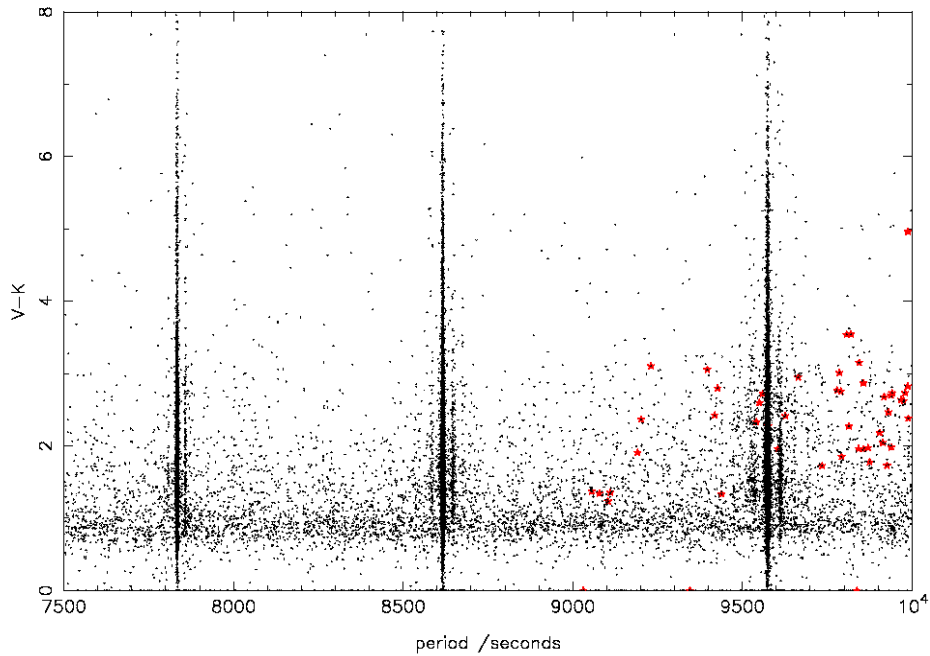}
\caption{The V--K colours of the variable stars found in the period range 7,500~sec to 10,000~sec. The candidate W UMa stars from Table 1 are shown as filled, larger symbols at {\em half} their proposed orbital period in each case. All are redder than the average colours of the rest of the sample. The structures seen amongst the other points are due to systematic noise giving rise to false periods close to 1/9, 1/10 and 1/11 of a sidereal day, with weaker structures at 1/9, 1/10 and 1/11 of a solar day.}
\end{center}
 \end{figure*}

\pagebreak

 \begin{figure*}[t]
\begin{center}
\includegraphics[scale=0.75,angle=0]{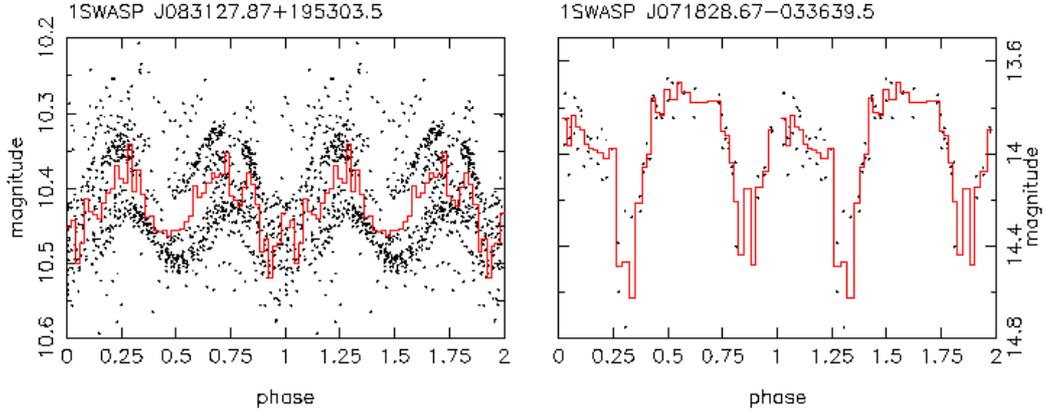}
\vspace{-15cm}
\caption{Two previously identified short period contact binaries which were not picked out by the current exercise owing to their relatively poorly sampled SuperWASP light curves. (a) The SuperWASP light curve of 1SWASP J083127.87+195303.5 folded at its orbital period of 0.2178~d. (b) The SuperWASP light curve of 1SWASP J071828.67--033639.5 folded at its orbital period of 0.2113~d.}
\end{center}
 \end{figure*}

 \begin{figure*}[t]
\begin{center}
\includegraphics[scale=0.5,angle=-90]{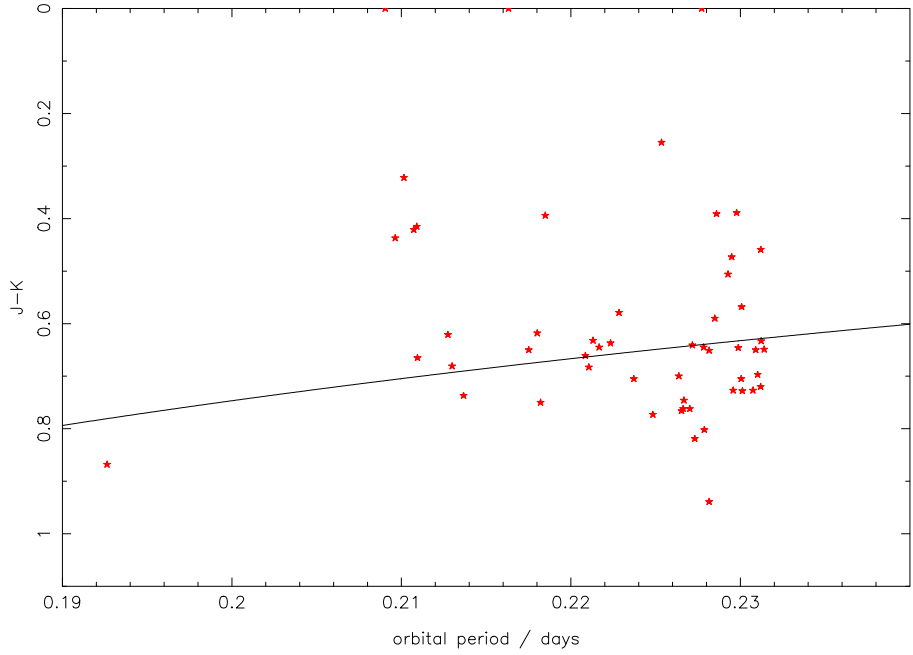}
\caption{The $J-K$ colours of the stars in our sample plotted against orbital period. Over-plotted is the period-colour relationship derived by Deb \& Singh (2010) for mostly longer period objects, shown in Equation 1.}
\end{center}
 \end{figure*}

 \pagebreak

\onecolumn


\begin{table}
\caption{Candidate SuperWASP short period eclipsing binaries}
\begin{tabular}{lccccl} \hline
 SuperWASP ID & Period/day & Maximum & Primary & Secondary & Other Name  \\ 
 1SWASP Jhhmmss.ss$\pm$ddmmss.s  &       &  SW V mag         & Depth     & Depth     &   \\ \hline
1SWASP J052036.84+030402.1  &  0.23140  &  12.35  &  0.27  &  0.27 & \\
1SWASP J220734.47+265528.6  &  0.23123  &  14.25  &  0.24  &  0.22 & \\
1SWASP J000437.82+033301.2  &  0.23119  &  14.53  &  0.42  &  0.40 & \\
1SWASP J170240.07+151123.5  &  0.23119  &  13.79  &  0.42  &  0.39 & ROTSE1 J170240.11+151122.7 \\
1SWASP J041655.13-492709.8  &  0.23102  &  15.04  &  0.43  &  0.35 & \\
1SWASP J051501.18-021948.7  &  0.23090  &  13.78  &  0.15  &  0.15 & \\
1SWASP J235333.60+455245.8  &  0.23074  &  14.75  &  0.38  &  0.25 & \\
1SWASP J232607.07-294130.7  &  0.23012  &  13.71  &  0.39  &  0.37 & \\
1SWASP J155822.10-025604.8  &  0.23008  &  13.77  &  0.18  &  0.18 & \\
1SWASP J150822.80-054236.9  &  0.23006  &  12.46  &  0.56  &  0.54 & \\
1SWASP J034439.97+030425.5  &  0.22988  &  14.25  &  0.20  &  0.17 & \\
1SWASP J161334.28-284706.7  &  0.22978  &  12.43  &  0.49  &  0.45 & \\
1SWASP J050904.45-074144.4  &  0.22958  &  13.29  &  0.64  &  0.59 & \\
1SWASP J111931.48-395048.2  &  0.22949  &  10.80  &  0.11  &  0.11 &  ASAS J111932--3950.8 \\
1SWASP J004050.63+071613.9  &  0.22928  &  12.25  &  0.10  &  0.09 & \\
1SWASP J210318.76+021002.2  &  0.22859  &  13.00  &  0.05  &  0.03 & \\
1SWASP J174310.98+432709.6  &  0.22853  &  13.19  &  0.44  &  0.40 & V1067 Her \\
1SWASP J214510.25-494401.1  &  0.22816  &  14.55  &  0.17  &  0.15 & \\
1SWASP J044132.96+440613.7  &  0.22815  &  14.19  &  0.39  &  0.33 & \\
1SWASP J180947.64+490255.0  &  0.22788  &  13.89  &  0.94  &  0.72 & V1104 Her \\
1SWASP J212454.61+203030.8  &  0.22783  &  14.35  &  0.15  &  0.13 & \\
1SWASP J093012.84+533859.6  &  0.22772  &   9.53  &  0.17  &  0.15 & \\
1SWASP J114929.22-423049.0  &  0.22731  &  14.19  &  0.70  &  0.55 & \\
1SWASP J120110.98-220210.8  &  0.22717  &  14.23  &  0.13  &  0.09 & \\
1SWASP J115557.80+072010.8  &  0.22701  &  15.18  &  0.33  &  0.27 & \\
1SWASP J030749.87-365201.7  &  0.22667  &  14.87  &  0.65  &  0.62 & \\
1SWASP J003033.05+574347.6  &  0.22663  &  14.89  &  0.68  &  0.27 & \\
1SWASP J160156.04+202821.6  &  0.22653  &  14.07  &  0.70  &  0.57 & \\
1SWASP J121906.35-240056.9  &  0.22637  &  15.25  &  0.63  &  0.47 & \\
1SWASP J092756.25-391119.2  &  0.22535  &  11.77  &  0.16  &  0.15 & \\
1SWASP J212808.86+151622.0  &  0.22484  &  14.49  &  0.47  &  0.34 & \\
1SWASP J173003.21+344509.4  &  0.22372  &  13.85  &  0.30  &  0.23 & \\
1SWASP J130920.49-340919.9  &  0.22284  &  13.91  &  0.19  &  0.16 & \\
1SWASP J040615.79-425002.3  &  0.22234  &  14.14  &  0.52  &  0.50 & \\
1SWASP J173828.46+111150.2  &  0.22168  &  13.80  &  0.24  &  0.24 & \\
1SWASP J183738.17+402427.2  &  0.22131  &  14.45  &  0.27  &  0.25 & \\
1SWASP J064501.21+342154.9  &  0.22105  &  14.11  &  0.35  &  0.35 & \\
1SWASP J074658.62+224448.5  &  0.22085  &  14.06  &  0.53  &  0.46 & \\
1SWASP J055418.43+442549.8  &  0.21825  &  12.73  &  0.36  &  0.29 & \\
1SWASP J224747.20-351849.3  &  0.21822  &  13.89  &  0.20  &  0.16 & \\
1SWASP J133105.91+121538.0  &  0.21801  &  10.36  &  0.22  &  0.18 & \\
1SWASP J080150.03+471433.8  &  0.21751  &  13.40  &  0.66  &  0.64 & \\
1SWASP J041120.40-230232.3  &  0.21633  &  13.95  &  0.41  &  0.36 & \\
1SWASP J234401.81-212229.1  &  0.21367  &  11.64  &  0.11  &  0.09 & \\
1SWASP J221058.82+251123.4  &  0.21299  &  15.13  &  0.18  &  0.16 & \\
1SWASP J195900.31-252723.1  &  0.21275  &  14.64  &  0.47  &  0.43 & \\
1SWASP J022727.03+115641.7  &  0.21095  &  14.95  &  0.28  &  0.24 & \\
1SWASP J115605.88-091300.5  &  0.21091  &  12.46  &  0.05  &  0.04 & \\
1SWASP J151652.90+004835.8  &  0.21073  &  13.97  &  0.10  &  0.10 & \\
1SWASP J092328.76+435044.8  &  0.21013  &  13.03  &  0.04  &  0.03 & \\
1SWASP J142312.63-222425.1  &  0.20964  &  13.19  &  0.06  &  0.05 & \\
1SWASP J210423.76+073140.4  &  0.20908  &  13.30  &  0.05  &  0.04 & \\
1SWASP J022050.85+332047.6  &  0.19264  &  13.13  &  0.31  &  0.19 & GSC 2314--0530\\ \hline 
\end{tabular}
\end{table}

\normalsize

\end{document}